\documentclass
[aps,prl,showpacs,twocolumn,amsmath,amssymb,a4paper,superscriptaddress]
{revtex4}
\usepackage{graphicx}
\begin{document}

\title{Electronic structure of the 
SrTiO$_3$/LaAlO$_3$ interface revealed by resonant soft x-ray 
scattering}

\author{H.~Wadati}
\email{wadati@phas.ubc.ca}
\affiliation{Department of Physics and Astronomy, 
University of British Columbia, Vancouver, 
British Columbia V6T 1Z1, Canada}

\author{D.~G.~Hawthorn}
\affiliation{Department of Physics and Astronomy, 
University of British Columbia, Vancouver, 
British Columbia V6T 1Z1, Canada}

\author{J.~Geck}
\affiliation{Department of Physics and Astronomy, 
University of British Columbia, Vancouver, 
British Columbia V6T 1Z1, Canada}

\author{T.~Higuchi}
\affiliation{Department of Advanced Materials Science, 
University of Tokyo, Kashiwa, Chiba 277-8561, Japan}

\author{Y.~Hikita}
\affiliation{Department of Advanced Materials Science, 
University of Tokyo, Kashiwa, Chiba 277-8561, Japan}

\author{H.~Y.~Hwang}
\affiliation{Department of Advanced Materials Science, 
University of Tokyo, Kashiwa, Chiba 277-8561, Japan}
\affiliation{Japan Science and Technology Agency, 
Kawaguchi 332-0012, Japan}

\author{S.-W.~Huang}
\affiliation{National Synchrotron Radiation 
Research Center, Hsinchu 30076, Taiwan}

\author{D.~J.~Huang}
\affiliation{National Synchrotron Radiation 
Research Center, Hsinchu 30076, Taiwan}

\author{H.-J.~Lin}
\affiliation{National Synchrotron Radiation 
Research Center, Hsinchu 30076, Taiwan}

\author{C.~Sch\"{u}\ss ler-Langeheine}
\affiliation{II. Physikalisches Institut, 
Universit\"{a}t zu K\"{o}ln, Z\"{u}lpicher Stra\ss e 77, 
D-50937 K\"{o}ln, Germany}

\author{H.-H.~Wu}
\affiliation{National Synchrotron Radiation 
Research Center, Hsinchu 30076, Taiwan}
\affiliation{II. Physikalisches Institut, 
Universit\"{a}t zu K\"{o}ln, Z\"{u}lpicher Stra\ss e 77, 
D-50937 K\"{o}ln, Germany}

\author{E.~Schierle}
\affiliation{Helmholtz-Zentrum Berlin f\"{u}r Materialien und Energie 
c/o BESSY, 
Albert-Einstein-Str.~15, D-12489 Berlin, Germany}

\author{E.~Weschke}
\affiliation{Helmholtz-Zentrum Berlin f\"{u}r Materialien und Energie 
c/o BESSY, 
Albert-Einstein-Str.~15, D-12489 Berlin, Germany}

\author{G.~A.~Sawatzky}
\affiliation{Department of Physics and Astronomy, 
University of British Columbia, Vancouver, 
British Columbia V6T 1Z1, Canada}

\date{\today}
\begin{abstract}
We investigated the electronic structure of the 
SrTiO$_3$/LaAlO$_3$ superlattice (SL) 
by resonant soft x-ray scattering. 
The (003) peak, which is forbidden for our 
``ideal'' SL structure, was observed 
at all photon energies, indicating reconstruction at 
the interface. 
From the peak position analyses taking into account 
the effects of refraction, we obtained evidence for 
electronic reconstruction of Ti $3d$ and O $2p$ states 
at the interface. 
From reflectivity analyses, we concluded 
that the AlO$_2$/LaO/TiO$_2$/SrO and 
the TiO$_2$/SrO/AlO$_2$/LaO interfaces 
are quite different, leading to 
highly asymmetric properties.  
\end{abstract}

\maketitle
%\section{Introduction}
Many oxide hetero-epitaxial devices confront 
the need to manage different possible interface atomic configurations, 
which can have important effects on the electronic structure 
at the most electrically sensitive regions of the device. 
For example, in (001) oriented manganite tunnel junctions or cuprate 
Josephson junctions, a perovskite such as SrTiO$_3$ (STO) 
is typically used as the insulating barrier. Because perovskites 
grow in unit cell blocks (a SrO/TiO$_2$ double layer for STO) 
in most growth techniques, the top and bottom interfaces 
across the barrier have different atomic terminations. 
Although most studies assume a symmetric barrier in these junctions 
neglecting this interface asymmetry, evidence is emerging that 
this is a crucial issue to understand and optimize \cite{hyamada}. 

One example of extremely anisotropic properties 
that arise as a function of interface termination is 
the interface between two band 
insulators STO and LaAlO$_3$ (LAO). This system is 
especially interesting due to the metallic 
conductivity \cite{Hwang2} and even 
superconductivity \cite{scSTOLAO} found at the interface. 
The electronic structure of this interface 
has been studied both 
experimentally \cite{nakagawa, huijben, Thiel, 
Brinkman, Herranz, Siemons, oxygen, intimage} 
and theoretically \cite{Pentcheva,mspark}, 
and there has been an intense debate about 
the origin of this metallicity, that is, whether it is due to 
oxygen vacancies (``extrinsic'') \cite{Siemons, oxygen} or due to the polar 
nature of the LAO structure \cite{nakagawa}, which could result in 
``electronic reconstruction'' as found in surfaces 
of polar materials by Hesper {\it et al} \cite{Hesper}. 
Photoemission spectroscopy has been recently used to observe 
the electronic structures of such 
interfaces directly \cite{takizawaLTOSTO, Hotta, WadatiLAOLVO, YM}, 
but it cannot be applied to the study of multilayers with 
several periods due to its surface sensitivity. 

In this study we investigated the electronic structure 
of the STO/LAO superlattice (SL) by resonant soft x-ray 
scattering \cite{AbbamonteRXS2}, 
which has recently been used to study 
LaMnO$_3$(LMO)/SrMnO$_3$(SMO) \cite{LMOSMO} 
and La$_2$CuO$_4$/La$_{1.64}$Sr$_{0.36}$CuO$_4$ \cite{LCOLSCO} SLs. 
Since x-ray scattering is a photon-in-photon-out process, 
resonant soft x-ray scattering is bulk-sensitive, 
and can be applied to insulators as well as metals. 
In this sense, this technique is complementary 
to photoemission, and 
well suited for studying the electronic structure 
of multilayers nondestructively. 
The (003) forbidden peak was used in Ref.~\cite{LMOSMO} 
to study the electronic structure of symmetric interfaces 
in LMO/SMO. Here we show that a similar experimental 
approach can be used to differentially probe the asymmetry 
of interfaces in ABO$_3$/A$^{\prime}$B$^{\prime}$O$_3$ SLs, 
where two distinctly different interfaces are formed. 
We introduce a model which is used to analyze 
the energy dependence of the scattering in terms of 
the known energy dependence of the absorption 
of the parent bulk compounds, by taking into account 
the effects of the photon energy dependent refraction, 
the  finite thickness of the SL, 
which removes the total extinction in otherwise forbidden (003) reflections, 
and the photon energy dependent absorption depth which again 
has an energy dependent finite size effect. 
From our study of 
the photon-energy-dependent (002) and (003) Bragg peak positions and 
the overall reflectivity spectra, 
we obtained evidence for electronic reconstruction and 
strong asymmetric properties at the two different 
STO/LAO interfaces. 

\begin{figure}
\begin{center}
\includegraphics[width=9cm]{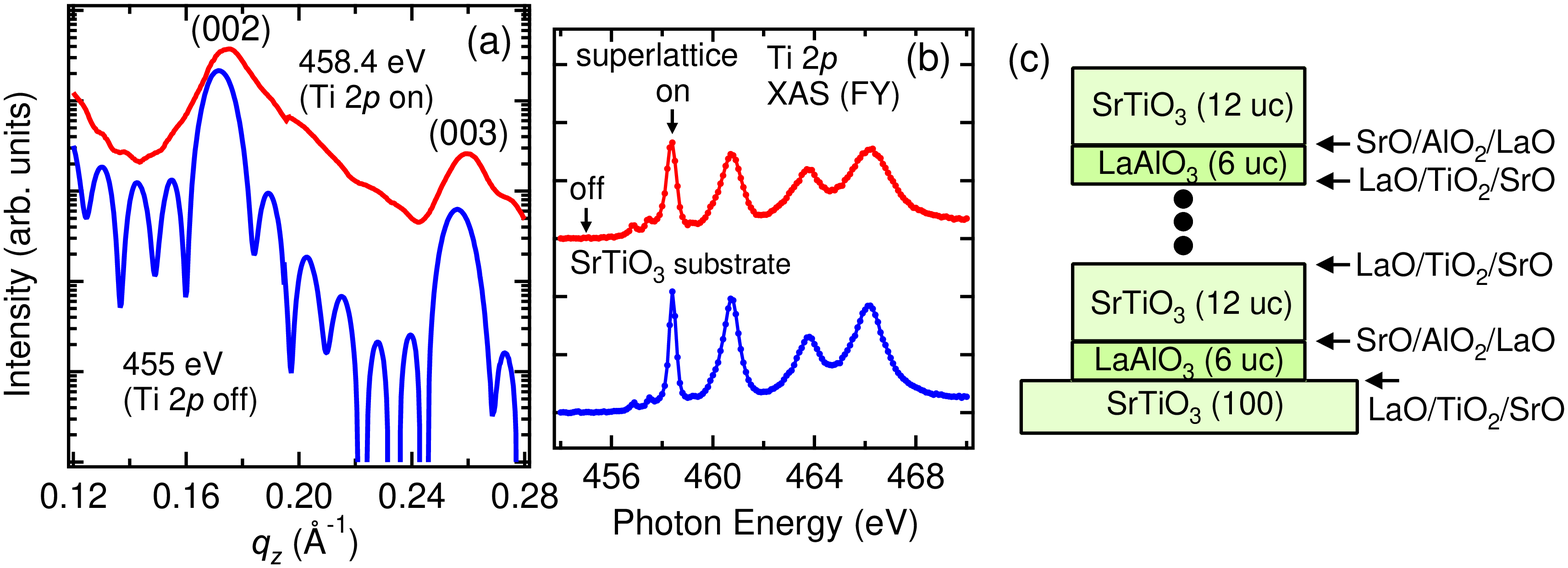}
\caption{(Color online) 
X-ray reflectivity and absorption spectra from 
the SrTiO$_3$/LaAlO$_3$ superlattice. 
(a) Reflectivity spectra measured at 458.4 eV 
(Ti $2p$ on-resonance) and 455 eV (Ti $2p$ 
off-resonance). 
(b) Ti $2p$ absorption spectra. 
(c) Schematic view of the SrTiO$_3$/LaAlO$_3$ superlattice.}
\label{fig1}
\end{center}
\end{figure}

%\section{Experiment}
The SL sample consisted of seven periods 
of 12 unit cells (uc) of STO and 6 uc of LAO. 
A schematic view of the fabricated SL is 
shown in Fig.~\ref{fig1} (c). 
The present samples were grown on a TiO$_2$-terminated 
STO (001) substrate \cite{Kawasaki} by 
pulsed laser deposition at an oxygen 
pressure of 1.0 $\times$ 10$^{-5}$ Torr and 
a substrate temperature of 1073 K. 
The lattice constant of one period of the SL 
was determined to be 73.45 $\mbox{\AA}$. 
%which is larger than the ideal value of 
%69.36 $\mbox{\AA}$ 
%(= $12\times 3.905$ (STO) + $6\times 3.75$ (LAO)), 
%where 3.75 $\mbox{\AA}$ is the $c$-axis lattice constant 
%for fully strained LAO thin films on STO substrates 
%(reflecting Poisson's ratio under tensile strain). 
The resonant soft x-ray scattering experiments were 
performed at the EPU beamline of NSRRC, Taiwan. 
The 
spectra were taken at 80 K. The incident light was 
polarized in the scattering plane ($\pi$ polarization) with the 
detector integrating over both final polarizations, i.e., 
both the $\pi \rightarrow \sigma$ and $\pi \rightarrow \pi$ 
scattering channels. We also measured 
x-ray absorption spectroscopy (XAS) 
spectra in the fluorescence-yield (FY) mode. 

%\section{Results and discussion}
Figure \ref{fig1} shows the x-ray 
reflectivity (a) and 
absorption (b) spectra measured 
in the energy region of Ti $2p$ absorption. 
Figure ~\ref{fig1} (b) shows that 
the XAS spectrum of the SL sample is 
almost the same as that of the STO substrate, 
which means that the formal valence of Ti is close to 
$4+$ in the SL. 
The reflectivity spectra in Fig.~\ref{fig1} 
(a) show finite-size Fresnel 
oscillations from the SL structures, 
and $(002)$, and $(003)$ Bragg peaks. 
The oscillations are clear at 
$455$ eV (Ti $2p$ off-resonance), but 
not evident at 458.4 eV (Ti $2p$ on-resonance). 
This is because the attenuation length of 
photons at $455$ eV is about 100 nm, which 
covers seven periods in the SL, but 
at $458.4$ eV it is only about 20 nm \cite{STOres}, 
which covers only three periods. 
Since the ratio of STO and LAO thicknesses 
are 2:1, the (003) peak is forbidden 
in infinitely thick and the zero absorption limit 
for samples of the ideal structure, just 
as in the case of the LMO/SMO 
SL studied in Ref.~\cite{LMOSMO}. In the case 
of LMO/SMO SL, the (003) peak was observed only 
on-resonance, but here it is observed in both 
on- and off-resonance. This is 
due to the strong asymmetry of the present SL, 
which contains two types of interfaces, 
the %$n$-type 
AlO$_2$/LaO/TiO$_2$/SrO and 
the %$p$-type 
TiO$_2$/SrO/AlO$_2$/LaO interfaces, 
as shown in Fig.~\ref{fig1} (c). 
In the LMO/SMO SL, there is only one kind of 
interface (SrO/MnO$_2$/LaO) because both LMO 
and SMO have MnO$_2$ layers. We now 
focus on (002), which is a normal 
Bragg peak, and (003), a forbidden 
interface-sensitive peak. 

\begin{figure}
\begin{center}
\includegraphics[width=8cm]{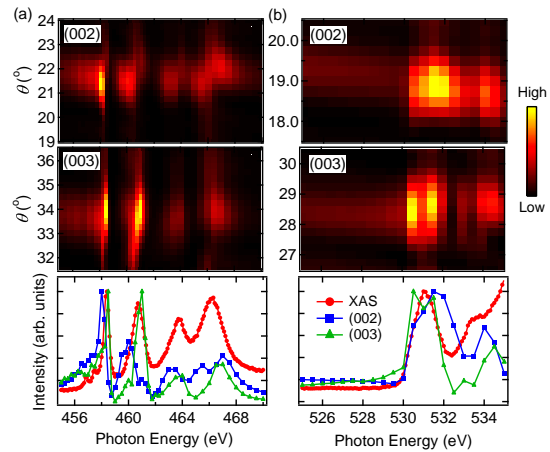}
\caption{(Color online) 
Photon-energy dependence of the (002) and (003) 
peaks near the Ti $2p$ (a) and O $1s$ (b) 
absorption edges. 
Top and middle panels show intensity maps 
of (002) and (003) regions, respectively. Here 
bright parts correspond to high intensities. 
Bottom panels show the (002) and 
(003) peak heights together with the 
XAS spectra.}
\label{fig2}
\end{center}
\end{figure}

Figure \ref{fig2} shows 
the photon-energy dependence of the (002) and (003) 
peaks near the Ti $2p$ (a) and O $1s$ (b) 
absorption edges. 
From the top and middle panels 
one can see that both the (002) and (003) peaks show 
resonant enhancement at these edges. 
The bottom panels show the (002) and 
(003) peak heights together with the 
XAS spectra. These peaks show enhancement 
where the Ti $2p$ or O $1s$ absorption is strong. 
In O $1s$ there is no evidence of pre-edge structures 
from states in the band gap. 
This is in sharp contrast to 
the results of the LMO/SMO SL \cite{LMOSMO}, where 
a pre-edge feature appears corresponding to 
states at the Fermi level. This 
difference can be explained by the difference of 
the mechanism of metallicity. 
In the case of LMO/SMO, metallic behavior is due to 
the hole-doping of insulating LMO, resulting in 
unoccupied states at the top of the valence band, 
whereas in the case of STO/LAO, 
metallicity is due to electron-doping of insulating STO so 
no new unoccupied states appear that would be probed 
by resonant soft x-ray scattering. 
We should note here that the energy-dependent 
behaviors of the (002) and (003) peaks are different, 
indicating a difference in the states at 
the interface from the bulk. 

We now analyze the peak positions in these 
absorption edges. In the normal Bragg's law 
$m\lambda = 2d\sin \theta$, 
where $m$ is an integer, and $d$ is the 
thickness of one unit cell of the SL. 
However, near the absorption edges, we must 
use the following modified Bragg's law \cite{att}, which 
takes into account the effects of refraction: 
\begin{equation}
 m\lambda = 2d\sin \theta 
\left(1-\frac{4\bar{\delta}d^2}{m^2\lambda^2}\right). 
\end{equation}
Here the refractive index $n$ is written as 
$n=1-\delta+i\beta$. 
For a system with ideally sharp interfaces 
between the two components, $\bar{\delta}$ is defined as 
the average of $\delta$'s of STO and LAO. 
\begin{equation}
\bar{\delta}=\frac{d_{\mathrm{STO}}\delta_{\mathrm{STO}}+
d_{\mathrm{STO}}\delta_{\mathrm{LAO}}}
{d_{\mathrm{STO}}+d_{\mathrm{LAO}}}\simeq
\frac{2\delta_{\mathrm{STO}}+\delta_{\mathrm{LAO}}}{3}, 
\end{equation}
where $d_{\mathrm{STO}}$ and $d_{\mathrm{LAO}}$ are 
the thickness of STO and LAO in one SL unit cell 
($d=d_{\mathrm{STO}}+d_{\mathrm{LAO}}$). 
$\delta$ and $\beta$ are related to the real and imaginary 
parts of the atomic scattering factors $f_1^0$ and 
$f_2^0$, respectively. The relationships are 
\begin{equation}
\delta=\frac{n_ar_e\lambda^2}{2\pi}f_1^0, \quad 
\beta=\frac{n_ar_e\lambda^2}{2\pi}f_2^0, 
\end{equation}
where $n_a$ is density of atoms and $r_e$ is 
the classical electron radius. 
Since $f_1^0$ cannot be determined experimentally, 
we use the relationship that $f_2^0$ is 
proportional to absorption and determined $f_2^0$ 
from the XAS spectra normalized to the values from 
Henke's table \cite{Henke}. 
$f_1^0$ is then obtained from the following 
Kramers-Kronig transformation 
\begin{equation} 
f_1^0(\omega)=Z-\frac{2}{\pi}P\int^{\infty}_{0}
\frac{uf_2^0(u)}{u^2-\omega^2}du, 
\end{equation}
where $Z$ is the atomic number and 
$P$ indicates 
the Cauchy principal value of integrals. 

\begin{figure}
\begin{center}
\includegraphics[width=9cm]{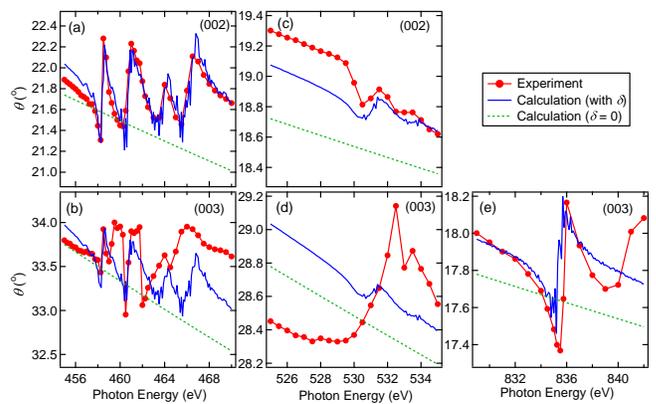}
\caption{(Color online) Peak-position analyses by considering the 
effects of refraction ($\delta$). 
(a) (002) at the Ti $2p$ edge. 
(b) (003) at the Ti $2p$ edge. 
(c) (002) at the O $1s$ edge. 
(d) (003) at the O $1s$ edge. 
(e) (003) at the La $3d$ edge.}
\label{fig5}
\end{center}
\end{figure}

Figure \ref{fig5} shows the analyses of 
the (002) and (003) peak positions by using 
the calculated $\delta$ and Eq.~(1). In the 
calculation, there is a large difference with finite 
$\delta$, which 
indicates that the effects of refraction are substantial 
in these absorption edges. From panel (a) and (c), 
one can see that at the (002) peak (normal Bragg peak) 
the agreement between experiment 
and calculation is good, which confirms the 
validity of our analyses. However, panels (b) and (d) 
show that at the (003) peak (interface-sensitive peak) 
the agreement is rather poor. We believe that this is 
due to the electronic reconstruction of Ti $3d$ 
and O $2p$ states at the interface, which changes 
the optical properties from those of the pure components. 
Eq.~(1) works well for the normal Bragg peak $(002)$, 
which is not sensitive to the interfaces, 
but for the interface-sensitive peak $(003)$, 
it does not provide a good description. 
This reveals that the underlying assumption 
of sharp interfaces is not appropriate and implies 
a different electronic structure and 
thereafter also a different energy-dependent 
dielectric constant from that assumed in Eq.~(1). 
From our experimental results alone, 
we cannot exclude the possibility that this is due to 
the existence of oxygen vacancies or disorder effects 
at the interface. 
However, the good agreement of the (002) peak assuming 
an ideal structure points toward electronic reconstruction. 
Panel (e) shows 
that at the La $3d$ edge the (003) peak follows 
the calculation, which indicates 
that there is not a serious interdiffusion of 
La atoms \cite{STOLAOdiff} at the interface. 

\begin{figure}
\begin{center}
\includegraphics[width=9cm]{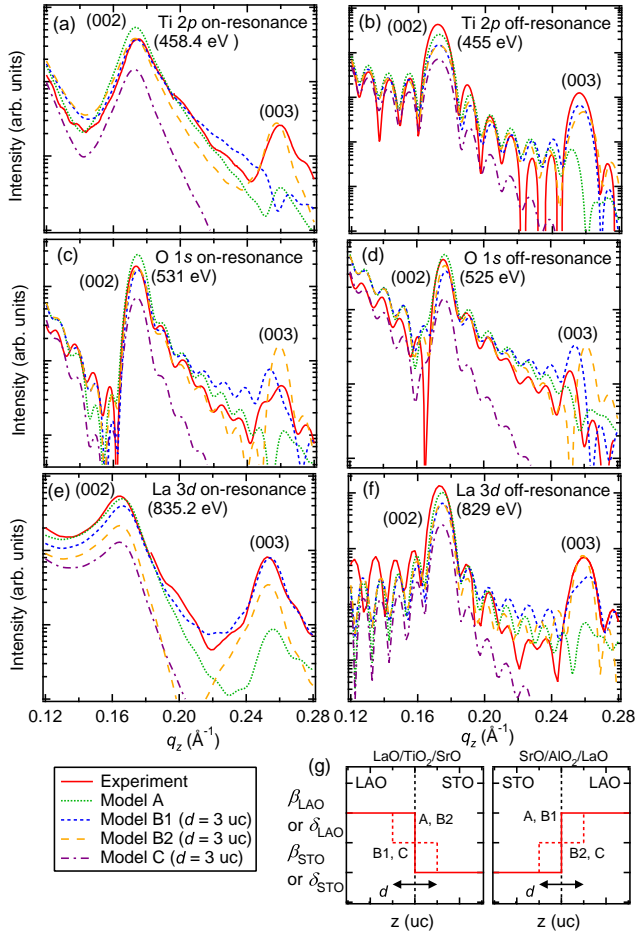}
\caption{(Color online) Comparison of the reflectivity spectra 
between experiment and calculation. 
(a) Ti $2p$ on-resonance. 
(b) Ti $2p$ off-resonance. 
(c) O $1s$ on-resonance. 
(d) O $1s$ off-resonance. 
(e) La $3d$ on-resonance. 
(f) La $3d$ off-resonance. 
(g) Models of our superlattice samples.}
\label{fig7}
\end{center}
\end{figure}

Next we analyzed the reflectivity spectra. 
The reflectivity can be simulated using the recursive 
Parratt's method \cite{parratt, under}. 
Here we considered four models as shown in 
Fig.~\ref{fig7} (g). Model A is the case where 
all interfaces are sharp. 
Model C is the case 
without any sharp interfaces with $\beta$ or $\delta$ at the interface 
taken to be the average of that of STO and LAO. 
These two models are considered 
as ``symmetric models'' because they do not consider 
the difference of AlO$_2$/LaO/TiO$_2$/SrO 
and TiO$_2$/SrO/AlO$_2$/LaO interfaces. 
Asymmetric models are models B1 and B2. 
These models include different interfaces 
as far as electronic roughness is concerned. 
In model B1, only the latter interfaces are sharp and 
in model B2, only the former are sharp. It was previously 
reported that metallic behavior is only observed at the 
former interface \cite{Hwang2} and also 
the former interfaces are atomically less sharp than 
the latter ones \cite{nakagawa}, seeming to support 
model B1. Motivated by these studies, we 
investigated which model can best describe 
the reflectivity spectra. 

Figure \ref{fig7} shows the comparison of the reflectivity spectra 
between experiment and calculation. From this figure one can see 
that the strong (003) peaks are not reproduced by models A and C, 
which demonstrates the necessity of the asymmetric models. 
In the two asymmetric models, model B1 reproduces 
the experimental results fairly well but 
for Ti $2p$ on-resonance, model B2 gives a 
better description of the experiment. 
Also we obtain the best fitting when 
the thickness of the interface ($d$) is taken to be about 3 uc. 
We should note here that in model A, with no reconstruction, 
the (003) peak is still present particularly at La 
on-resonance. This is due to the short penetration depth 
of the x-rays at resonance, providing imperfect 
extinction as described before. 
From these results, we conclude that our SL is a highly 
asymmetric system with two different types of interfaces, 
and the thickness of the interface is about 3 uc. 

%\section{conclusion}
In summary, we investigated the electronic structures of the STO/LAO SL by 
resonant soft x-ray scattering. From reflectivity measurements, 
we observed oscillations due to superlattice structures. 
The forbidden (003) peak was observed even at off-resonance, in 
sharp contrast to the case of LMO/SMO \cite{LMOSMO}. 
Both (002) and (003) Bragg peaks show resonant enhancement 
at the Ti $2p$, O $1s$, and La $3d$ absorption edges. 
From the peak position analyses taking into account 
the effects of refraction, the behavior of (002) and 
(003) at the La $3d$ absorption 
could be reproduced by calculations but that of (003) 
at the Ti $2p$ and O $1s$ absorption edges could not be 
reproduced by the model which does not take into account 
changes in the atomic scattering factors due to electronic 
reconstruction. These results point to electronic 
reconstruction of Ti $3d$ and O $2p$ at the interface.  
From reflectivity 
analyses, we found evidence for highly asymmetric 
properties of the STO/LAO SL, which means that 
the AlO$_2$/LaO/TiO$_2$/SrO and 
the TiO$_2$/SrO/AlO$_2$/LaO interfaces are 
quite different. The thickness of 
the interface was determined to be about 3 uc. 

%\section{acknowledgments}
The authors would like to thank 
I. Elfimov and A. Fujimori for informative discussions. 
H. W. acknowledges financial support from the 
Japan Society for the Promotion of Science. 
J. G. acknowledges the support by DFG. 
C. S.-L. acknowledges the support by the DFG 
through SFB 608. 
This work was made possible by financial 
support from the Canadian funding Agencies 
NSERC, CRC, CIFAR, and CFI. 

\bibliography{LVO1tex}
\end{document}